\documentclass[epjCONF, onecolumn]{svjour}
\usepackage{graphics}
\usepackage[varg]{txfonts} 
\usepackage[latin1]{inputenc}
\session-title{Assembling the puzzle of the Milky Way}
\begin{document}
\title{The chemical evolution of the Galactic Bulge seen through micro-lensing events}
\author{Sofia Feltzing\inst{1}\fnmsep\thanks{\email{sofia@astro.lu.se}} \and Thomas Bensby\inst{1}  \and  Jorge Mel\'{e}ndez\inst{2} \and Daniel Ad\'en\inst{1} \and  Martin Asplund\inst{3} \and Andy Gould\inst{4} \and  Jennifer Johnson\inst{4} \and Sara Lucatello\inst{5} \and  Avishay Gal-Yam\inst{6}  }
\institute{Lund Observatory, Box 43, SE-221 00 Lund, Sweden \and Universidade de S\~{a}o Paulo, IAG, Rua do Mat\~{a}o 1226,   Cidade Universit\'{a}ria, S\~{a}o Paulo 05508-900, Brazil \and Research School of Astronomy and Astrophysics, Australian National University, Cotter Rd., Weston, ACT 2611, Australia \and Department of Astronomy, Ohio State University, 140 W. 18th   Avenue, Columbus,   OH 43210, USA \and Osservatorio Astronomico di Padova, Vicolo dell'Osservatorio 5, 35122 Padua, Italy \and Benoziyo Center for Astrophysics, Faculty of Physics, The Weizmann Institute of Science, Rehovot 76100, Israel}
\abstract{ Galactic bulges are central to understanding galaxy
  formation and evolution.  Here we report on recent studies using
  micro-lensing events to obtain spectra of  high resolution and
  moderately high signal-to-noise ratios of dwarf stars in the Galactic
  bulge. Normally this is not feasible  for the faint turn-off stars in the
  Galactic bulge, but micro-lensing offers this possibility. Elemental
  abundance trends in the Galactic bulge as traced by dwarf stars are
  very similar to those seen for dwarf stars in the solar
  neighbourhood. We discuss the implications of the ages and
  metallicity distribution function derived for the micro-lensed
  dwarf stars in the Galactic bulge.  } 
\maketitle
\section{Introduction}
\label{intro}
\vspace{-0.3cm}

Galactic bulges are emerging as inherently complex features in spiral
galaxies. Numerous studies have shown them to have several
spatial structures overlaying each other. The Milky Way is no
different -- over the last decades our view of the stellar content,
gas and stellar dynamics in the inner few kpc of our own galaxy has
developed significantly. If the spatial structures are uniquely 
related to dynamical and chemical features is a very actively 
studied field. 

New surveys such as VISTA Variables in The Via Lactea (VVV) public
survey (described in \cite{refsaito}) are looking deeper and deeper in
to this intriguing Galactic component. Several outstanding questions
remain un-resolved including the true shape of the metallicity
distribution function (MDF) and how the MDF is connected to different
spatial and kinematical structures. A recent discussion of these
issues can be found in \cite{carine}.  Other questions concern the star
formation history of the bulge and the presence or absence of age
spreads as, for example, traced by asymptotic giant stars
\cite{vanloon}.

The history of a stellar population is imprinted in its stars.  The
elemental abundances in the atmospheres of stars often remain
unperturbed over time and act as time-capsules showing the mixture of
elements present in the gas from which the stars formed. This is in
particular true for dwarf stars and their spectra, even for metal-rich
 stars, are fairly straightforward to analyze and are the best
tracers of galactic chemical evolution \cite{edvardsson}. However,
dwarf stars in the Galactic bulge are too faint to be observed under
normal circumstances ($V$= 19 -- 20, as seen in for example
colour-magnitude diagrams obtained with the HST, \cite{feltzing}).
The chemical history of the bulge has therefore mainly been studied
using intrinsically bright giant stars. Results based on giant spectra
are not trivial to interpret as evolutionary processes erase some of the
abundance information and the cool atmospheres of giants, rich in
molecules, are difficult to analyse, see discussion in
 \cite{fulbright}.  IR spectroscopy of bulge
giants has recently become feasible but is still limited by the very
restricted wavelength coverage on existing spectrographs. A recent
example is given by the CRIRES spectra analysed in
\cite{ryde}. However, the underlying assumption that giants accurately
represent all stars has not yet been rigorously tested, as discussed in
\cite{taylor}.  Therefore, a metallicity distribution function based
on red giant star may not reflect the original distribution.  Physical
processes within the red giant stars can also lead to the erasure of
some of the original abundance signatures.  This is in particular the
case for C, N, and Li.  These abundances are eventually altered in all
red giants and in some stars O, Na, Mg, and Al are also altered as discussed in
\cite{kraft}. This means that a true study of the star formation
history in the Galactic bulge requires the study of dwarf stars.
Furthermore, the precision achieved in dwarfs is better than in
giants, allowing us to look for any substructure that may be present
in the bulge population.  Finally, for dwarf stars close to the
turn-off point or on the sub-giant branch it is possible to derive
individual ages. These give us a unique insight in to the age
structure of the Galactic bulge.

 Micro-lensing offers the unique opportunity to
  observe dwarf stars in the bulge. When the star
  is lensed by a foreground object its magnitude can
  increase more than 5 magnitudes making it possible to obtain a
  spectrum of high resolution and S/N so that a standard abundance
  analysis can be done \cite{bensby2011}.

  The observed micro-lensing events discussed in this contribution
  and their analysis are fully documented in \cite{bensby2010} and
  \cite{bensby2011}. Issues concerning limb-darkening are further
  developed in \cite{johnson}.

\section{Micro-lensed dwarf stars in the Galactic Bulge - discussion}
\label{sec:1}
\vspace{-0.3cm}

\begin{figure}
\resizebox{0.7\columnwidth}{!}{%
\includegraphics{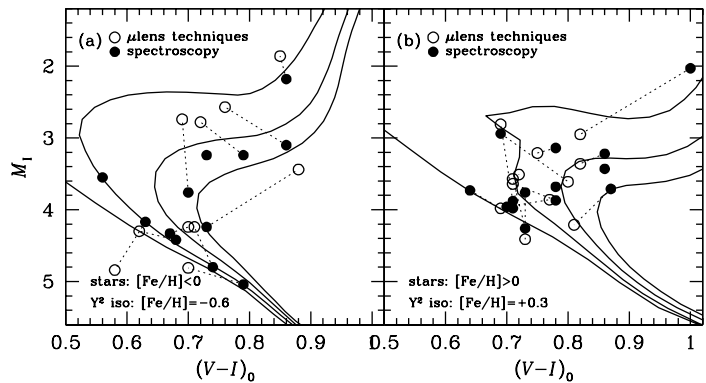}}
\caption{Colour-magnitude diagrams for the micro-lensed dwarf
  stars. The colours and magnitudes are determined using micro-lensing
  techniques ($\circ$) and spectroscopy ($\bullet$). {\bf a} Shows the
  stars with [Fe/H]$<0$ and {\bf b} the stars with [Fe/H]$>$0.  Each
  colour-magnitude diagram also show representative $Y^2$ isochrones
  for 1, 5, 10, and 15\,Gyr, from \cite{demarque}.  For each star we
  connect the result based on micro-lensing techniques with that based
  on spectroscopy using a dotted line. 4 stars have no values from
  micro-lensing techniques (see \cite{bensby2011}). }
\label{feltzingfig:1}       
\end{figure}

\subsection{Abundance trends}
\vspace{-0.3cm}
We find that the elemental abundance trends in the Galactic bulge as
traced by the micro-lensed dwarf stars is very similar, if not
identical, to that found in the solar neighbourhood for dwarf stars
with kinematics typical of the thick disk \cite{bensby2011}. Recent
studies of red giant stars have also shown great similarities between the
local thick disk giants and the giants in the Galactic bulge
\cite{alvesbrito}. Thus it appears that the earlier results where red
giant stars in the Galactic bulge showed large $\alpha$-enhancements
also at solar and even super-solar metallicities must be ascribed to
the difficulty in analysing optical spectra of metal-rich red giants
(see also discussion in \cite{bensby2010}).
\vspace{-0.3cm}

\subsection{MDF and IMF}
\label{subsec:1}
\vspace{-0.3cm}
In our two papers \cite{bensby2010} and \cite{bensby2011} we compare
the MDF based on the small number of micro-lensed dwarf stars
available with the then best MDF based on spectroscopy of red giant
stars as analysed in \cite{zoccali2008}. We found a significant
difference with the dwarf stars showing a bi-modal MDF. A recent
re-analysis of the spectra of these red giant stars has changed the
situation somewhat by making the MDF based on giant stars somewhat
more bi-modal. This re-analysis is presented in \cite{hill}. A KS-test
between that result and the MDF based on micro-lensed dwarf stars
(including only those presented by us in \cite{bensby2011}) is
inconclusive.  The MDFs could be drawn from the same population or
not.

The most recent update of our MDF based on micro-lensed dwarf
stars, in total 37 stars (end of September 2011), still 
shows a bi-modal MDF and with an ever increasing fraction 
of metal-rich stars. 

An interesting implication of the true shape of the MDF concerns the
Initial Mass Function (IMF). The origin of the slope of the IMF is
much debated (for a short, recent introduction to the debate see
\cite{oey}).  Many processes leads to roughly the same slope and there
is not much evidence that metallicity influences the shape or slope of
the IMF, a review can be found in \cite{bastian}.  Recent work on the
IMF in the Galactic bulge has resulted in interesting conclusions.
The peak of the MDF depends on the slope of the IMF.  The chemical
evolution model of the Galactic bulge by \cite{ballero} is based on
the photometric MDF of giant stars by \cite{zoccali2003} and on the
spectroscopic MDF of giant stars by \cite{fulbright}. In
\cite{ballero} an IMF  much
flatter than in the solar neighbourhood is found, i.e., an IMF skewed
towards high mass stars.

A more recent model used the spectroscopic MDF, still using the
original results from \cite{zoccali2008}, also requires a flat IMF to
reproduce the peak of the MDF as derived from the red giant stars
(\cite{cescutti}). However, the MDF based on micro-lensed dwarf stars
persistently shows two well-defined peaks, one that can be associated
with a metal-poor old bulge, and another with super-solar
metallicities that can be associated with a younger population
(compare also the discussion of the connection between the stellar
kinematics and their metallicities presented in \cite{carine}). Hence,
the IMF no longer has to be flat to explain a single peaked solar
metallicity MDF that was made in 0.5 Gyr, as in these recent models
(\cite{cescutti}). To reproduce the bi-modal nature of the bulge MDF,
probably a normal IMF can be used for the metal-poor bulge, while
contributions from type Ia SN might explain the younger metal-rich
peak. However, more detailed models are needed before firm conclusions
can be drawn.
\vspace{-0.3cm}

\subsection{Ages}
\label{subsec:2}
\vspace{-0.3cm}

\begin{figure}
\begin{center}
\resizebox{0.55\columnwidth}{!}{%
\includegraphics{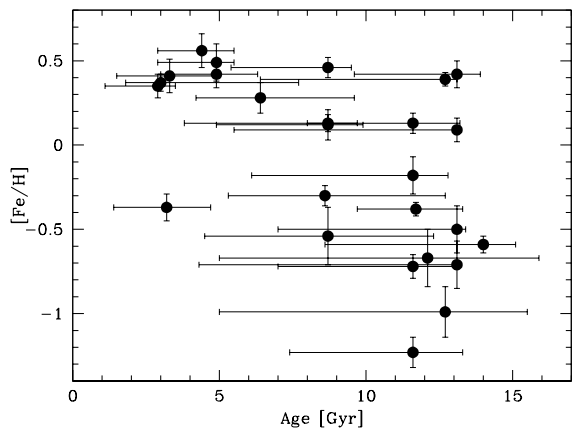} }
\end{center}
\caption{The ages and metallicities of the 26 dwarf stars 
as derived in \cite{bensby2011}.}
\label{feltzingfig:2}       
\end{figure}

Dwarfs near the turn-off are unique as we can get stellar ages for
them. Figures\,\ref{feltzingfig:1} and \ref{feltzingfig:2} show a
summary of results of our determination of stellar parameters and ages
for stars presented in \cite{bensby2011}. In Fig.\,\ref{feltzingfig:1}
we split the stars according to metallicity. We show two different
values of $M_{\rm I}$ and $(V-I)_0$ for each star. One is based on the
effective temperature and surface gravity derived from the stellar
spectra alone (spectroscopic values). The micro-lensing technique
relates the magnitude of the star to that of the red clump stars in
the same field.  The advantages and draw-backs of each technique is
discussed in our recent paper, \cite{bensby2011}. The thing to take
away from the left-hand panels in Fig.\,\ref{feltzingfig:1} is that
regardless of which technique is used, stars with sub-solar
metallicities essentially trace an old turn-off, while stars with
super-solar metallicities show a wider range of ages. This is still
true when the latest events are included (Bensby et al. 2012 in
prep.).

A surprising result from the micro-lensed dwarf stars is the presence
of a large age spread among the most metal-rich stars.  This result
might appear unexpected given the large amount of evidence based on
deep CMDs that show a red and faint turn-off, most often interpreted
as the result of a uniquely old and metal-rich stellar population (as
shown in numerous studies, including \cite{holtzman}, \cite{ortolani},
\cite{feltzing} \cite{zoccali2003}, \cite{clarkson}). However, there
is evidence from AGB stars of an intermediate age population in the
Galactic bulge. This has been seen at least in three 
independent studies (\cite{vanloon}, \cite{cole}, \cite{uttenthaler}).

Based on ISOGAL and DENIS data of the inner 10$^\circ$ of
the Galactic bulge \cite{vanloon} find a few hundred asymptotic giant
branch stars, which is consistent with their inferences from the near
infrared CMDs and
\cite{uttenthaler} find evidence for Tc in a sub-sample of their
C-stars, indicative of third dredge up and a minimum stellar mass of
1.5\,M$_{\odot}$ which implies an upper age limit of 3\,Gyr.  Our data
for micro-lensed dwarfs appear to confirm the existence of such an
intermediate age population in the inner kpcs.
\vspace{-0.3cm}

\section{Summary and outlook}
\vspace{-0.3cm}

So far we have presented elemental abundances and ages for 26
micro-lensed dwarf stars (\cite{bensby2010} and
\cite{bensby2011}). They show that dwarf stars in the Galactic bulge
share the elemental abundance trends with the thick disk in the solar
neighbourhood and have an MDF that is bi-modal. This is also true for
the most recent observations (Bensby et al. 2012 in prep.). Surprisingly, we
find, among the stars with super-solar metallicities, a wide range of
ages. This remains to be better explained, but we note that AGB stars
and variable stars (Miras) present in the Galactic bulge also point to
a sub-population with an intermediate age. We note that this is, on 
the surface, contradictory to the red and faint turn-offs seen in
all CMDs of the Bulge. However, a smaller intermediate age, metal-rich stellar
population can most likely still be accommodated. Detailed modelling of
this and larger samples of both dwarf and giant stars in the 
Galactic bulge covering wider areas of the bulge are needed to fully
understand the connection between the various spatial structures in the 
bulge and the MDF.
\vspace{-0.5cm}

\end{document}